# EVALUATE THE PERFORMANCE OF OPTICAL CROSS CONNECT BASED ON FIBER BRAGG GRATING UNDER DIFFERENT BIT RATE

Bobby Barua
Assistant Professor, Department of EEE, Ahsanullah University of Science and Technology, Bangladesh
bobby@aust.edu

## ABSTRACT

*In wavelength division multiplexed (WDM) optical networks, intraband and/or interband crosstalk plays a major role in limiting practical implementation of an OXC subsystem. Again FIBER BRAGG GRATINGS (FBGs), due to their properties, are emerging as very important elements for both the optical fiber communication and sensing fields. The crosstalk sources are related to the different individual components of the OXC's. In this paper, an analytical model is established to evaluate the performance of an FBG-OC-based optical cross connect and proposed a novel scheme of transmission Fiber Bragg Grating for different wavelength separation and determine the BER and power penalty with optical wavelength division multiplexed cross-connect topologies at different bit rate.*

## 1. INTRODUCTION

OPTICAL wavelength division multiplexing (WDM) networks are very promising due to their large bandwidth, their large flexibility and the possibility to upgrade the existing optical fiber networks to WDM networks [1-6]. WDM has already been introduced in commercial systems. Crosstalk analyses of OXCs presented so far are generally focused on conventional OXCs [7, 8]. All-optical cross connects (OXC), however, have not yet been used for the routing of the signals in any of these commercial systems. Several OXC topologies have been presented in the literature, but their use has so far been limited to field trials, usually with a small number of input–output fibers and/or wavelength channels. The fact, that in practical systems many signals and wavelength channels could influence each other and cause significant crosstalk in the optical cross connect, has probably prevented the use of OXC's in commercial systems. The crosstalk levels in OXC configurations presented so far are generally so high that they give rise to a significant signal degradation and to an increased bit error probability. Because of the complexity of an OXC, different sources of crosstalk exist, which makes it difficult to optimize the component parameters for minimum total crosstalk. The crosstalk sources are related to the different individual components of the OXC's. The crosstalk levels of the four topologies are compared in function of the number of cascaded OXC's [9]. For evaluation of intra-band crosstalk in an FBG-OC-Based optical cross connect an analytical model is established to evaluate intra-band crosstalk performance and a novel scheme with improved intra-band crosstalk performance [10-11]. Incoherent crosstalk is defined as the case in which the beat term can be neglected (e.g., when the wavelengths are different). The case in which the beat term cannot be neglected is called coherent crosstalk. This crosstalk occurs in a WDM network if channels with the same nominal carrier frequency are combined. The impact of incoherent crosstalk on signal transmission performance has been studied in[12], but the details of incoherent crosstalk have not been analyzed, as we will show that such crosstalk may also be a coherent combination of crosstalk contributions and then cause much higher noise power[13]. This paper studied the statistical impact of coherent and incoherent crosstalk in an OXC and in optical networks. In addition, the analysis is extending of crosstalk in a FBG.

## 2. SYSTEM BLOCK DIAGRAM

A general structure for such WDM Transmission system with optical cross connects are shown in Fig.1. Here, N nodes of a cross connect are shown in a two sided model with the transmit side on the left and the receive side on the right. As shown in the figure, Data comes from any source. Each transmitted signal carries a number of unique wavelengths which correspond to the destination data, here several channel multiplexed and go optical cross connect.

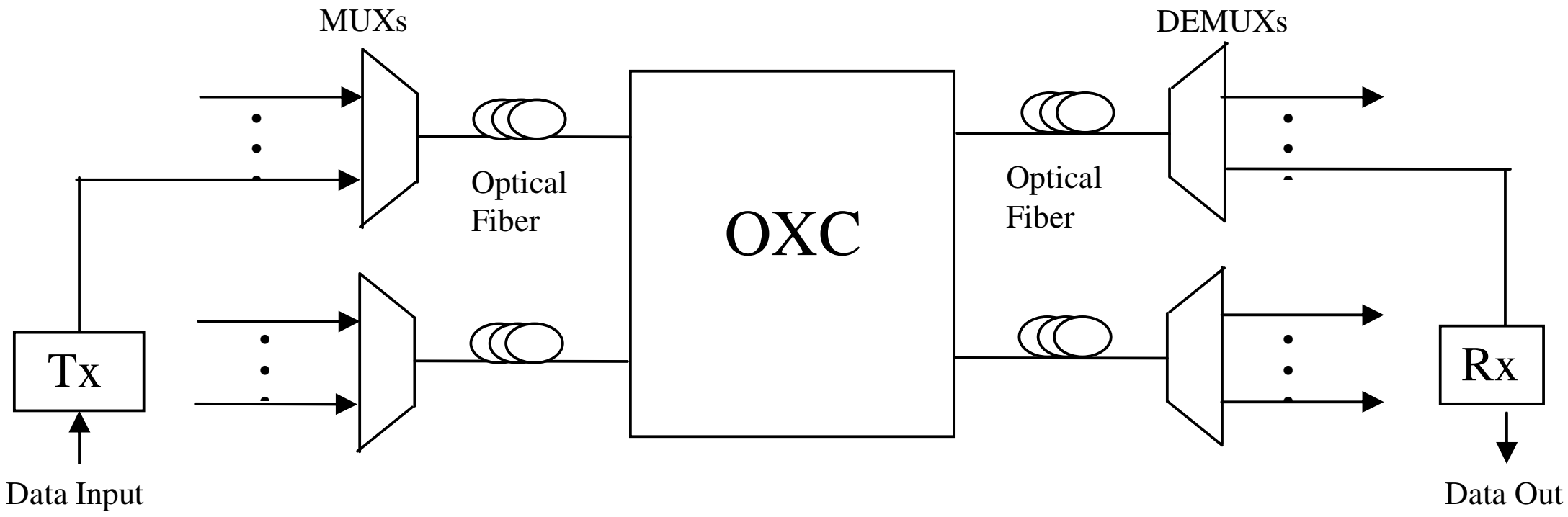

**Fig.1 :** Block diagram of an Optical WDM Transmission system with an Optical Cross connect (OXC)

All the optical signals are wavelength-multiplexed and are transmitted over a single mode fiber to the network hub. At the hub wavelength de multiplexers separate the signals from each incoming fiber. All channels intended for a given destination are passively re arranged, and since they were allocated different wavelength, they can be wavelength multiplexed and sent towards the destination on a single fiber. At the receiver of each node, the different channels are wavelength de multiplexed.

## 3. PRINCIPLE OF OPERATION OF AN FIBER BRAGG GRATING (FBG)

The FBG in its most basic form consists of a periodic modulation of the refractive index along a single mode fiber. The FBG is a distributed reflector, acting as a narrowband reflection filter. The maximum reflectivity of the FBG occurs at a wavelength that matches the Bragg condition:

$$\lambda_B = 2n_{eff}\Lambda \qquad (1)$$

where the Bragg grating wavelength $\lambda_B$ is the free-space center wavelength of the input light that will be back-reflected from the FBG, $n_{eff}$ is the effective refractive index of the fiber core, and $\Lambda$ is the FBG period. When an ultrasonic wave impinges on an optical fiber with a FBG (immersed in water), the refractive index of the fiber and the FBG period are modulated due to ultra sound induced mechanical strain in the fiber through the elastooptic effect. For the case of a single mode fiber and plane ultrasonic wave at normal incidence the changes in the physical length L and refractive index n of the fiber can be written as a function of the pressure ΔP as:

$$\frac{\Delta\Lambda}{L} = -\frac{(1-2V)\Delta P}{E} \quad (2)$$

$$\frac{\Delta n}{n} = \frac{n^2 \Delta P}{2E}(1-2V)(2p_{12}+p_{11}) \quad (3)$$

where E and v are the Young's modulus and Poisson's ratio of the fiber, ΔP is the pressure variation that in our case is caused by ultrasound, and p12 and p11 are components of the strain-optic tensor for the fiber material. Given ΔL/L = ΔΛ/Λ, the spectral shift of the FBG as a function of pressure becomes:

$$\Delta\lambda_B = \lambda_B \left[ -\frac{(1-2V)}{E} + \frac{n^2}{2E}(1-2V)(2p_{12}+p_{11}) \right] \Delta P \quad (4)$$

The last expression shows that the spectral shift of the FBG is directly proportional to the acoustic pressure ΔP. Xu et al. have reported $\Delta\lambda_B$ / ΔP to be 0.003 nm/MPa over hydrostatic pressure range of 70 MPa for a FBG with peak reflectivity of 80%, and bandwidth of 0.7 nm at a Bragg wavelength of 1533 nm. It should be noted that this approximation is valid only for the case of normal incidence of low-frequency acoustic waves with wavelengths significantly larger than the diameter of the optical fiber. If the acoustic wavelength is comparable to the diameter of the optical fiber and/or the ultrasonic wave impinges on the fiber at an angle, the stress distribution along the FBG will no longer be uniform. In such cases, non-uniform changes of the FBG pitch as well as other factors such as axially guided ultrasonic waves described in the following sections, must be taken into account.

## 4. FIBER BRAG GRATING (FBG) BASED OXC

The configuration of OXC using FBGs and six-port OCs . There are two sets of tunable FBGs ($A_1$ and $B_1$ ) and four sets of fixed FBGs ($A_2$ , $A_3$ ,$B_2$ , and $B_3$ ) whose function will be explained subsequently. In the figure, we use $A_i$ to denote the i-th set of FBGs whose Bragg wavelengths are associated with odd-number wavelengths $\{\lambda_{2n-1}\}$(n=1,2,3,……..m) of group A and $B_i$ to present the i-th set of FBGs whose Bragg wavelengths are associated with even-number wavelengths $\{\lambda_{2n}\}$(n=1,2,3,……..m) of group B. The OCs used here are fully rotatable. A fully rotatable OC (ROC) can route the input signal from the last port back to the first port. The WDM signals of group A at wavelengths $\{\lambda_{2n-1}\}$ entering at port 1 of ROC1 are reflected by Set $A_2$ of fixed FBGs whose Bragg wavelengths are fixed and matched with the incoming WDM channels. These WDM signals will go through port 3 of ROC1, port 3 of ROC2 and encounter the tunable FBGs of Set $A_1$ . Note that a tunable FBG with Bragg wavelength only needs to be tuned to its adjacent Bragg wavelength $\lambda_{2n}$ or $\lambda_{2n-2}$ , and hence, the tuning range required is equal to the channel spacing. By tuning the Bragg wavelength of each tunable FBG, the incoming WDM signals will be either passed through or reflected back. A reflected signal will be coupled back to port 3 of ROC1 and exit at port 4 of ROC1, whose switching state in the OXC is denoted as a "bar" state. On the other hand, a through signal will be coupled into port 3 of ROC2 and exit at port 4 of ROC2, whose switching state in the OXC is denoted as a "cross" state. Similarly, for the WDM signals of group B at wavelength $\{\lambda_{2n}\}$that are launched into port 1 of ROC2, we can achieve the same independent switching by tuning the tunable FBGs of Set $B_1$ .

## 5. ANALYSIS OF CROSSTALK IN A FBG

The total E-field, including the homowavelength crosstalk

$$E_M(t) = (E_M + \sum_{n \neq m}^{N} E_N)\exp(-iW_M t) \qquad W_M = \frac{2\pi c}{\lambda_M} \tag{5}$$

The received photocurrent, including of the beat term

$$I_P(t) = RP_M(t) + 2R\sum_{N \neq M}^{N-1}\sqrt{P_M(t)P_N(t)}\cos(\varphi_M(t) - \phi_N(t))$$
$$+ 2R\sum_{N,J=2(N \succ J)}^{N-1}\sqrt{P_N(t)P_J(t)}\cos(\phi_N(t) - \phi_J(t)) + R\sum_{N \neq M}^{N-1} P_N \tag{6}$$

The mean and variance of the photo-current are given as:

$$E\left|I_P(t)\right| = R\int_{-\infty}^{\infty}(P_m(t) + S_{CT}(t) + P_{ASE}) \bullet h(t-\tau)d\tau$$

$$\mathrm{var}\left|I_P(t)\right| = \sigma_{TH}^{\ 2} + R\int_{-\infty}^{\infty}(P_m(t) + S_{CT}(t) + P_{ASE}) \bullet h^2(t-\tau)d\tau$$
$$+ R^2 \cdot \frac{1}{\beta}\left[\frac{P_{ASE}^{\ 2}}{2}\int_{-\infty}^{\infty}h^2(t+\tau)d\tau + P_{ASE}\int_{-\infty}^{\infty}(P_M(t) + S_{CT}(t)) \bullet h^2(t-\tau)d\tau\right] \tag{7}$$

The above enable the numerical modeling of penalty

$$\sigma^2_{TOTAL} = \sigma^2_{TH} + \sigma^2_{SHOT} + \sigma^2_{ASE-ASE} + \sigma^2_{ASE-Signal} + \sigma^2_{ASE-Crosstal} \tag{8}$$

BER in presence of Crosstalk

$$\sigma^2_O = \sigma^2_{thermal} + \sigma^2_{Shot} \tag{9}$$

$$BER = \frac{1}{4} \cdot erfc\left(\frac{1}{\sqrt{2}} \cdot \frac{R_D P_S}{2\sqrt{\sigma_O^{\ 2} + (N-1)X_{SW} R_D^{\ 2} P_S^{\ 2}}}\right) \tag{10}$$

$X_{sw}$ = Optical switch crosstalk
N = number of cross-points

## 6. ANALYSIS OF CROSSTALK WITH OXC BASED A FBG

The analytical equations for the FBG-OC crossconnect are depicted in this section. Here in this analysis the signal power is defined by $P_i^j$ , where i denotes the wavelength channel and j denotes the number of fibers. $J_0$ designates the fiber which contain the channel under study and $i_0$ designates the wavelength under study. The input power of the channel under study is defined by $P_{i0}^{j0}$ and the optical power at the output of the first stage is defined by out io $P_{io1}^{out}$

$$P_{io1}^{out} = P_{io}^{jo} + \left(P_{io}^{j} X_{FG} + P_{io}^{jo} X_{OC}\right) - 2 \sqrt{P_{io}^{j} P_{io}^{jo}} \sqrt{X_{FG}} - 2\sqrt{P_{io}^{jo} P_{io}^{j}} \sqrt{X_{OC}} - 2 \sqrt{P_{io}^{jo} P_{io}^{j}} \sqrt{X_{FG} X_{OC}} \tag{11}$$

Among the five contributions of this equation the first term is the input signal, the second and third terms express the crosstalk contributions and the last three are beating terms. Here it is assumed that for the worst case the sign of the beat terms are negative and the amplitude of the beat terms are maximum. Here $X_{OC}$ is the optical circulator crosstalk and $X_{FG}$ is the fiber Bragg grating (FBG) crosstalk which is given by,

$$X_{FG} = 10\log_{10}\left(1 - R_{FG}\right) \tag{12}$$

where R $_{FG}$ is the FBG reflectivity. $P^{j}io$ is the wavelength channel power at another fiber j that carries a wavelength $i_o$ . Let $P_{io1}^{out(ref)}$ is the output of wavelength channel $i_0$ when the FBGOC cross connect carries only wavelength channel $i_0$ (when there is no crosstalk). Then the crosstalk can be expressed as-

$$crosstalk = \frac{P_{io}^{out(ref)} - P_{io}^{out}}{P_{io}^{out(ref)}} \tag{13}$$

## 7. Crosstalk Analysis :

The electrical field of the main signal after passing through the BOXc can be written as

$$E_o(t) = \sqrt{P_o \chi_o} \exp\left\{ j\left[ wt + \phi_o(t) \right] \right\} e_o \tag{14}$$

$\chi_0$ = Transmission loss for the main signal and $e_o$ = unit polarization vector of the signal
For $Y_1$ coherent crosstalk fields

$$E_k(t) = \sqrt{P_o \chi_k \chi_o} \exp\left\{ j\left[ wt + \phi_k(t) \right] \right\} e_k \;;\; k = 1,2,3................Y_1 \tag{15}$$

For $Y_2$ incoherent crosstalk fields

$$E_1(t) = \sqrt{P_o \chi_1 \chi_o} \exp\left\{ j\left[ wt + \phi_1(t) \right] \right\} e_1 \;;\; k = 1,2,3................Y_2 \tag{16}$$

Where $\chi_k$ and $\chi_1$ are the power ratio of each coherent and incoherent field

$$\chi_k = \frac{P_{cc}(\Delta\lambda)}{P_s} \quad and \quad \chi_1 = \frac{P_{nc}(\Delta\lambda)}{P_s}$$

$e_k$ and $e_l$ are the unit polarization vector of the coherent and incoherent crosstalk field.

$$E_{tot} = E_o(t) + \sum_{k=1}^{Y_1} E_k(t) + \sum_{l=1}^{Y_2} E_l(t) \tag{17}$$

Because $E_o(t)$ and $E_k(t)$ arise from the same light source $\lambda_i$, $\phi o$ and $\phi_k$ are correlated with each other. Thus, Based on the parallelogram law, there must be a field $E_1$, which is the vector addition of $E_o(t)$ and $\sum_{k=1}^{Y_1} E_k(t)$. Similarly, since all the incoherent crosstalk items arise from the same light sources λ`i, there is a field $E_2$, which is the vector additions of $\sum_{l=1}^{Y_2} E_l(t)$. If, $b_0(t)$ = binary bit sequences of the light sources $\lambda_i$ and $b_1(t)$ = Binary bit sequences of the light sources $\lambda_i$, and $P_1$= optical power of the fields $E_1$ ; $P_2$= optical power of the fields $E_2$, then, the total power of the main signal with all coherent and incoherent cross talk items received by the photo detector cab be :

$$P_{tot} = \left| E_o(t) + \sum_{k=1}^{Y_1} E_k(t) + \sum_{l=1}^{Y_2} E_l(t) \right|^2 \approx b_o(t)P_1 + 2b_o(t)b_1(t)\sqrt{P_1P_2}\cos(\Delta\phi) \tag{18}$$

$\Delta\phi$ = phase difference between $E_1$ and $E_2$

The first term in the right hand side is the power of the main signal combined with all coherent crosstalk items; the second term is the beating noise between fields $E_1$ and $E_2$. Here, we assume that the optical propagation delay difference between the main signal and a coherent crosstalk field is much less than the time duration of one bit. Therefore, coherent crosstalk does not cause noise but changes the power of the main signal. $P_1$ and $P_2$ are the optical power of the fields $E_1$ and $E_2$, respectively.

$$P_1 = \left| E_o(t) + \sum_{k=1}^{Y_1} E_k(t) \right|^2$$
$$= P_o\chi_o \left\{ 1 + \sum_{k=1}^{Y_1} \chi_k + 2\sum_{k=1}^{Y_1} \sqrt{\chi_k}\cos(\Delta\phi_k) + 2\sum_{k=1}^{Y_1}\sum_{l=1}^{Y_1} \sqrt{\chi_k\chi_l}\cos(\Delta\phi_k - \Delta\phi_l) \right\} \tag{19}$$

$$P_2 = \left| \sum_{l=1}^{Y_2} E_l(t) \right|^2 = P_o\chi_o \left\{ 1 + \sum_{l=1}^{Y_2} \chi_l + 2\sum_{l=1}^{Y_2}\sum_{m=1}^{l-1} \sqrt{\chi_l\chi_m}\cos(\Delta\phi_l - \Delta\phi_m) \right\} \tag{20}$$

where $\Delta\varphi_k$, $\Delta\varphi_l$ and $\Delta\varphi_m$ are the phase offset of coherent and incoherent crosstalk items related to the phase of the main signal.

## 8. Analysis of Bit Error Rate (BER)

The bit-error-rate (BER) of a WDM transmission system consisting of a bidirectional FBG-OC based OXC can be expressed as:

$$BER_{worstcase} = \frac{1}{8}\left[ Q\left( \frac{1}{\sqrt{2}} \frac{i_1 + i_{CTO} - i_D}{\sigma_{1-0}} \right) + Q\left( \frac{1}{\sqrt{2}} \frac{i_D - i_{CTO} - i_0}{\sigma_{0-0}} \right) + Q\left( \frac{1}{\sqrt{2}} \frac{i_1 + i_{CT1} - i_D}{\sigma_{1-1}} \right) + Q\left( \frac{1}{\sqrt{2}} \frac{i_D - i_{CT1} - i_0}{\sigma_{0-1}} \right) \right] \quad (21)$$

Where $Q(\chi) = (1/\sqrt{2\pi} \int_{\chi}^{\infty} e^{-(u^2/2)} du)$ and $i_1 = \Re_D\ P_{i0}^{j0}$ is the photocurrent for transmitted bit "1" where $\Re_D$ is the photo detector responsively and $i_0$ is the photocurrent for transmitted bit "0". The decision threshold current, $i_D$ can be express as :

$$i_D = \frac{\sigma_{0-1} i_1 + \sigma_{1-1} i_0}{\sigma_{0-1} + \sigma_{1-1}} \quad (22)$$

In the following equation $\sigma^2_{1-0}$ is the noise variance when signal bit "1" is interfered by crosstalk due to bit "0", $\sigma^2_{0-0}$ is the noise variance when signal bit "0" is interfered by crosstalk due to bit "0", $\sigma^2_{1-1}$ is the noise variance when signal bit "1" is interfered by crosstalk due to bit "1"and $\sigma^2_{0-1}$ is the noise variance when signal bit "0" is interfered by crosstalk due to bit "1". These four noise variances can be defined as:

$$\sigma^2_{0-1} = \sigma^2_{th} + 2e\Re_D \left( P_{i0}^{j0} - P_{CT0} \right) B \quad (23)$$

$$\sigma^2_{0-0} = \sigma^2_{th} + 2e\Re_D P_{CT0} B \quad (24)$$

$$\sigma^2_{1-1} = \sigma^2_{th} + 2e\Re_D \left( P_{i0}^{j0} - P_{CT1} \right) B \quad (25)$$

$$\sigma^2_{0-1} = \sigma^2_{th} + 2e\Re_D P_{CT1} B \quad (26)$$

---

e denotes the electronic charge and $\sigma^2_{th}$ is the variance of the thermal noise in the detector with a temperature of 300K. $\sigma^2_{th}$ can be expressed as :

$$\sigma^2_{th} = \frac{4KTB}{R_L} \quad (27)$$

K denotes the Boltzman constant, T denotes the receiver temperature, B denotes the electrical bandwidth of the receiver and $R_L$ denote the receiver front end load. $P_{CT0}$ and $P_{CT1}$ represent the crosstalk power due to bit "0" and "1" respectively and can be express as :

$$P_{CTO} = -P_{i0}^{out0} \quad (28)$$

$$P_{CT1} = P_{i0}^{j0} - P_{i0}^{out1} \quad (29)$$

## 9. RESULTS AND DISCUSSION

Following the theoretical analysis presented in previous sections, the performance results of an optical WDM transmission link with optical cross-connect based on Fiber Brag Grating (FBG) are evaluated taking into considerations the effect of crosstalk due to OXC. The results are presented in terms of amount of crosstalk due to OXC and system BER a given number of input wavelength with several values of channel separation. The power penalty suffered by the system due to OXC induced crosstalk at a given BER is also determined.

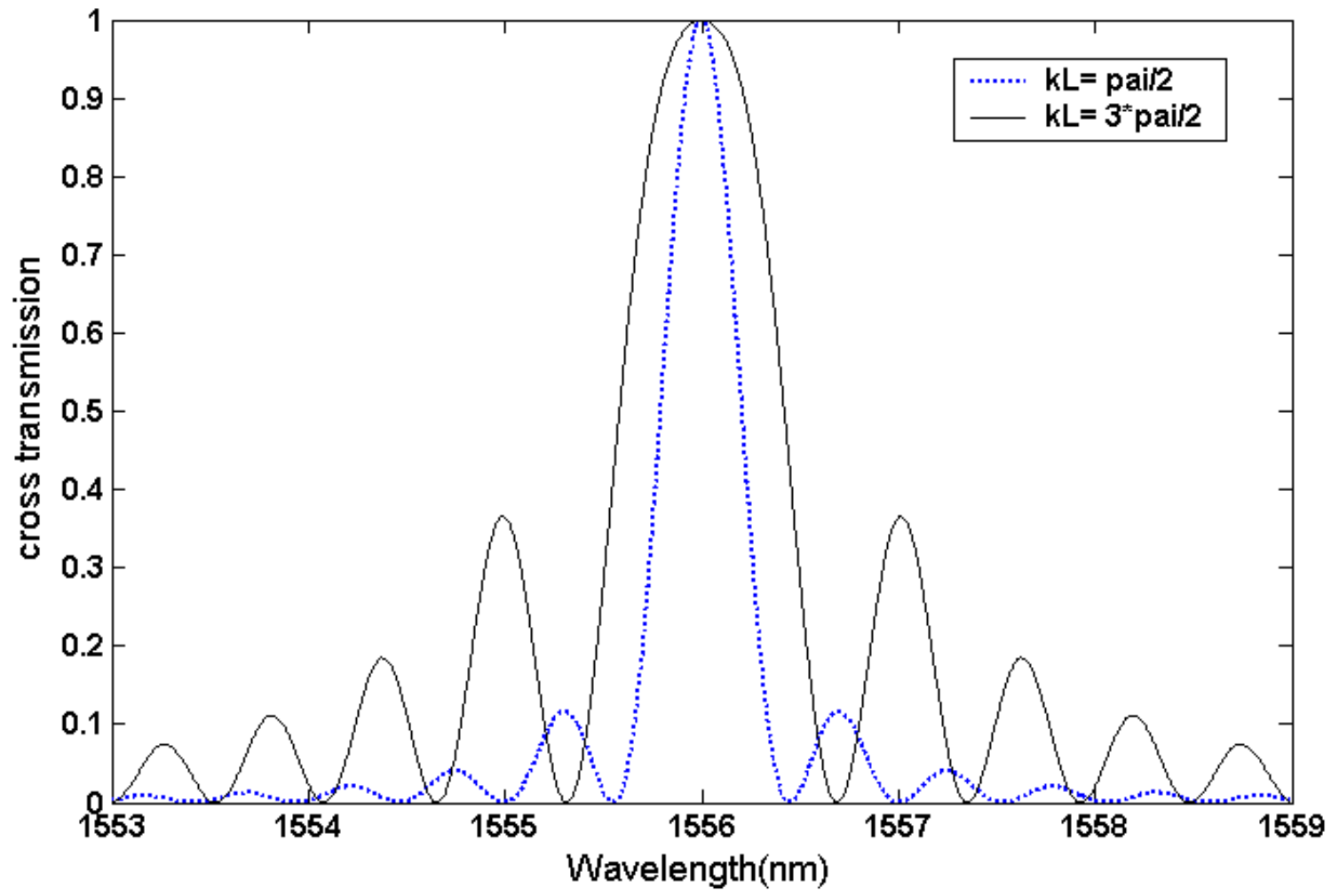

**Fig.2:** Calculated cross transmission t_ through uniform transmission gratings with $kL = \pi/2$ (dashed line) and $kL = 3\pi/2$ (solid line).

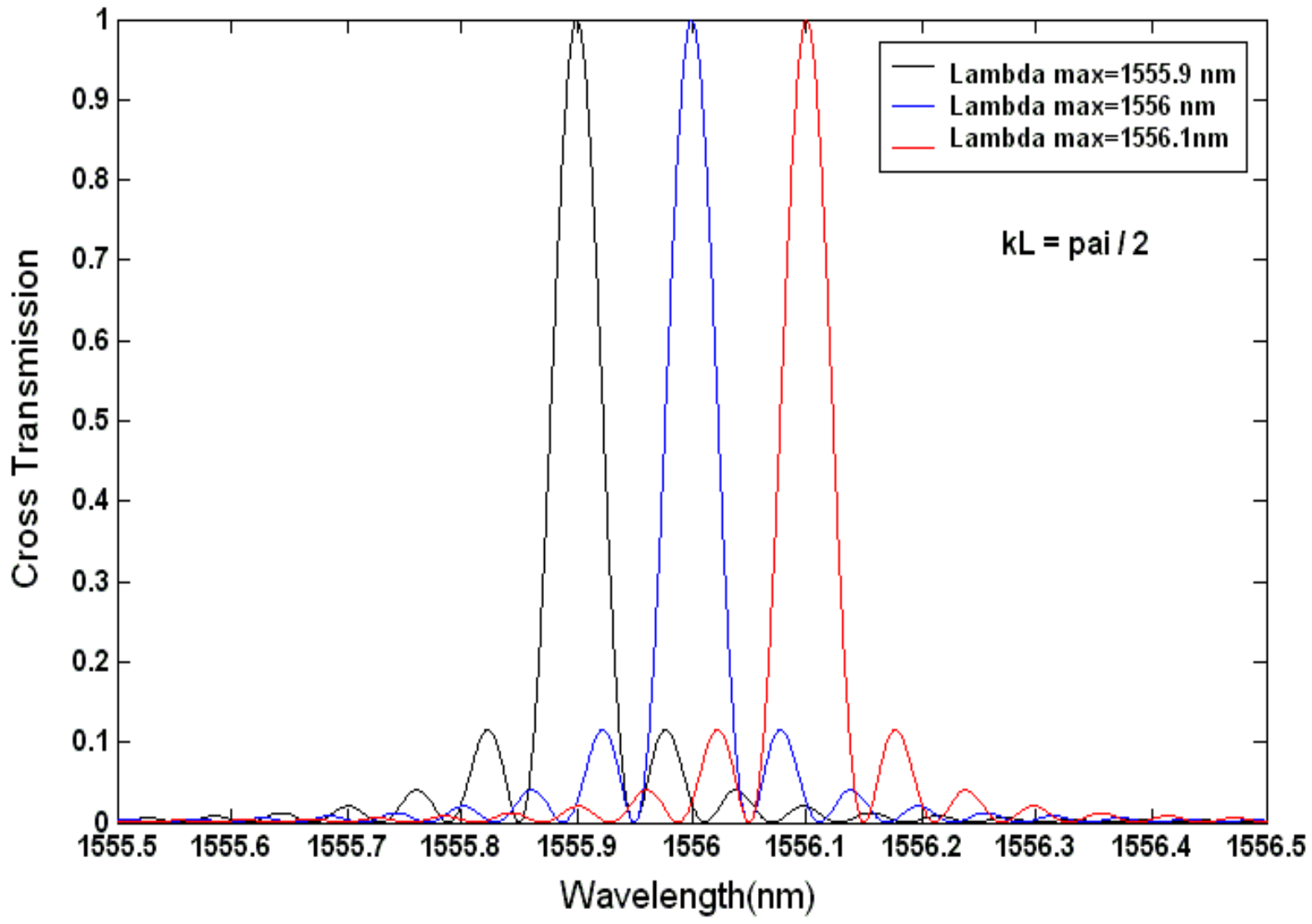

**Fig. 3 :** Plots of three channel crosstalk for $kL = \pi/2$ .

The transmittance of the Fiber Bragg Grating considered in the OXC is evaluated and is shown in Fig.2 as a function of input optical wavelength for two values of KL, via $\pi/2$ and $3\pi/2$. The figure depicts wavelength of the FBG is 1556 nm with 3dB optical pass band for $kL = \pi/2$ and $3\pi/2$ respectively. The transmittance of an OXC with three WDM channels with wavelengths1555.9 nm,1556nm and 1556.1 nm is shown in Fig.3. The figure depicts the crosstalk introduced by the OXC which transmitting the three wavelength channels due to long frequency of the transmittance function. It is notified that maximum amount of crosstalk is introduced in the middle channel. Similar plots of the cross transmition of the OXC is shown in Fig.4 and Fig.5 for number of wavelength N=5 and 7 respectively. As the number of channels are increased, the amount of crosstalk introduced became higher.

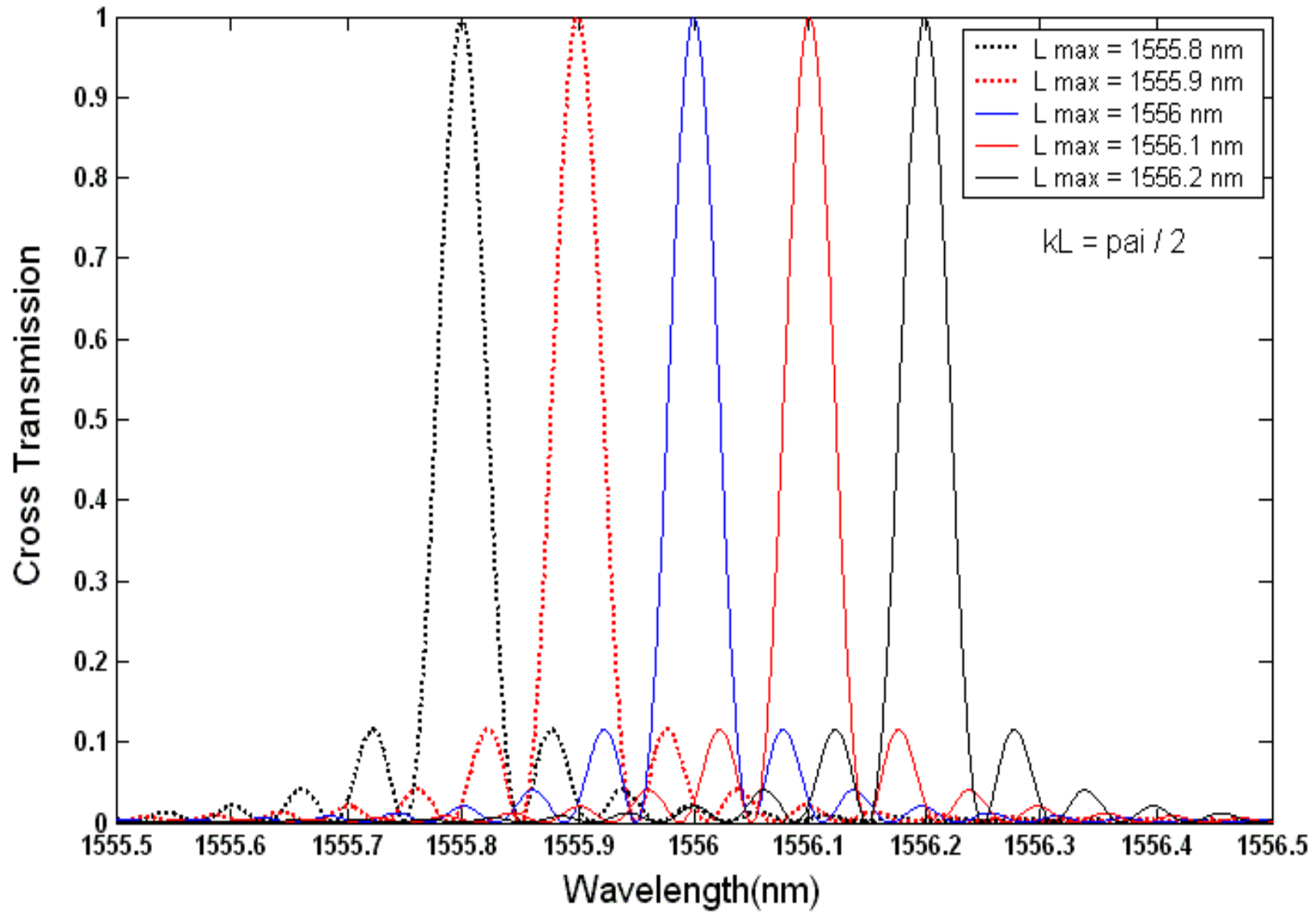

**Fig.4 :** Plots of five channel crosstalk for kL = $\pi$ /2 .

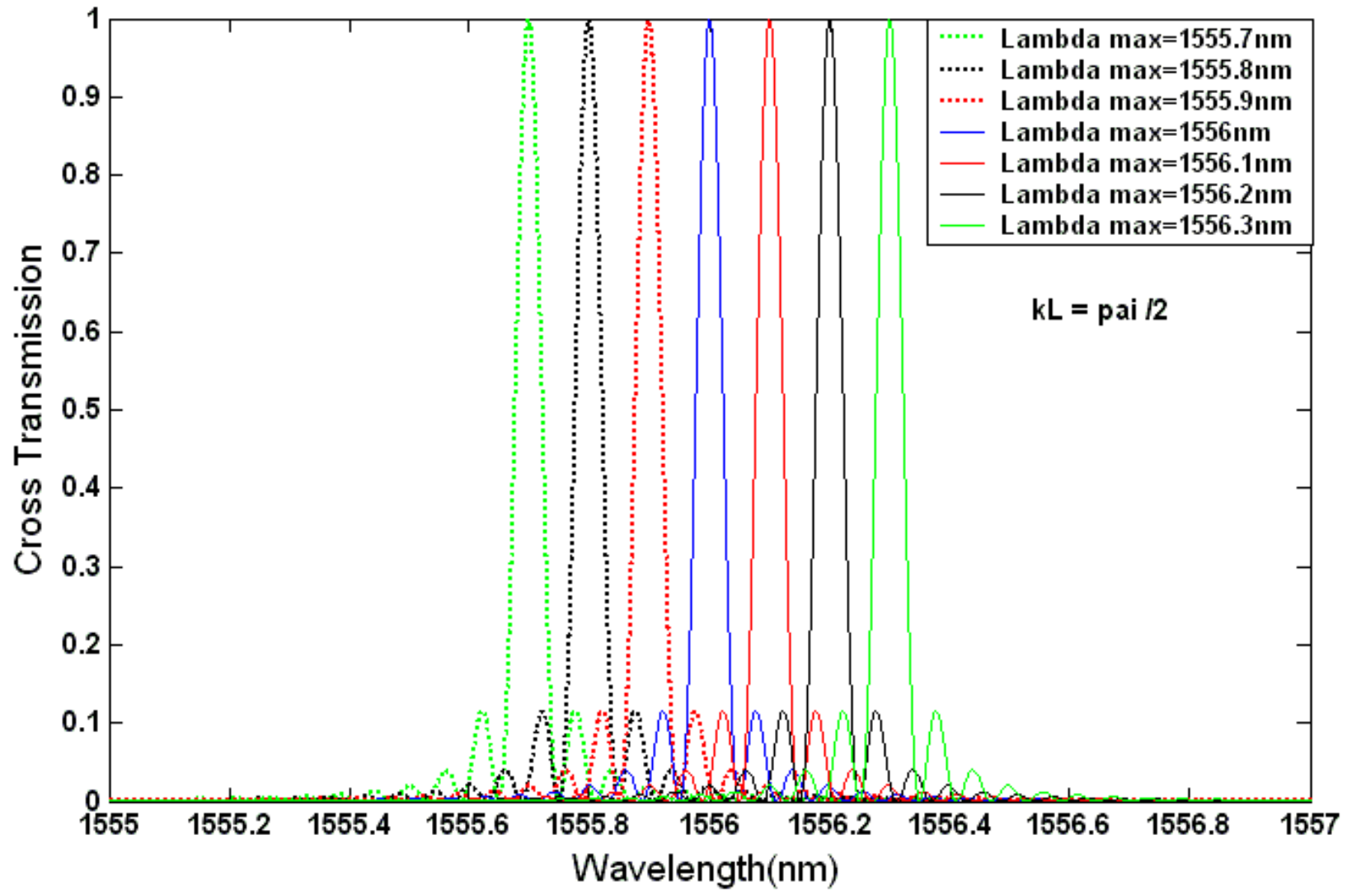

**Fig.5:** Plots of seven channel crosstalk for kL = $\pi$ /2 .

The relative amount of crosstalk introduced in the middle channel is evaluated by integrating the transmition the channel bandwidth. The plots of crosstalk due to OXC verses channel separation among the WDM channels are depicted in Fig.6 for number of channel N=3,5 ,7. with kL =$\pi$/2**.** It is noticed that the optical signal power is significantly high at lower value of channel separation and exponentially decreases with increase in channel separation. For example, the relative crosstalk level is –20dB corresponding to $\Delta\lambda$ =0.1nm and is reduced to –40dB when $\Delta\lambda$ is increased to 1.0nm for N=7, for N=3, the crosstalk level is slightly lower. The bit error rate (BER) performance results of a WDM system with an OXC based a FGB and operation at a bit rate of 1Gbps with relative crosstalk level Pc/Pin as a parameter. BER performance of the system without crosstalk is also depicted for comparison. It is reveled that there is a deterioration in the BER performance due to crosstalk which even cause an BER floor at higher value of Pc/Pin. For example, there occurs a BER floor at around $10^{-8}$ for Pc/Pin = -21.34 dB which correspond to N=3.

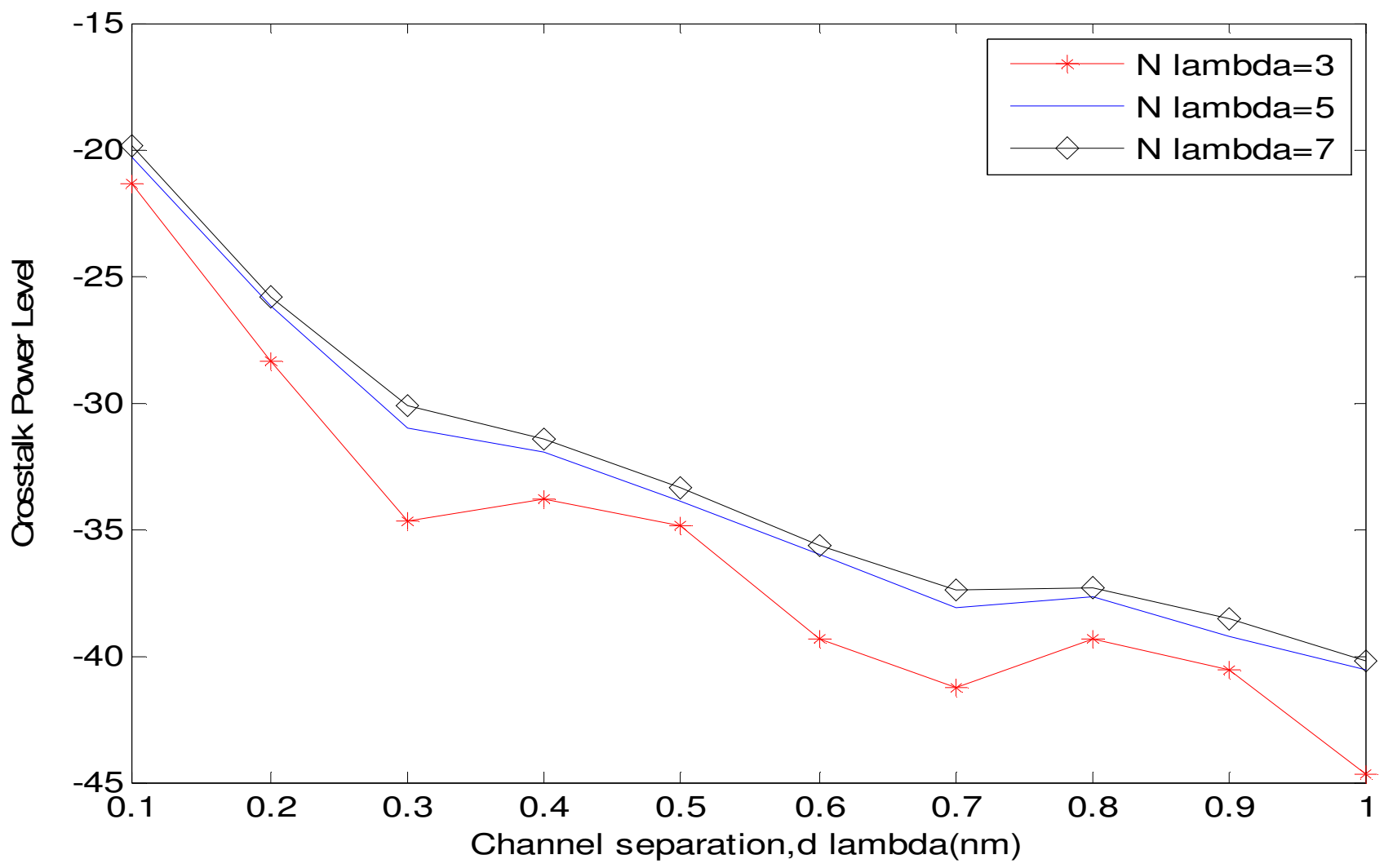

**Fig.6 :** Plots of crosstalk power versus channel separation for $kL = \pi/2$ .

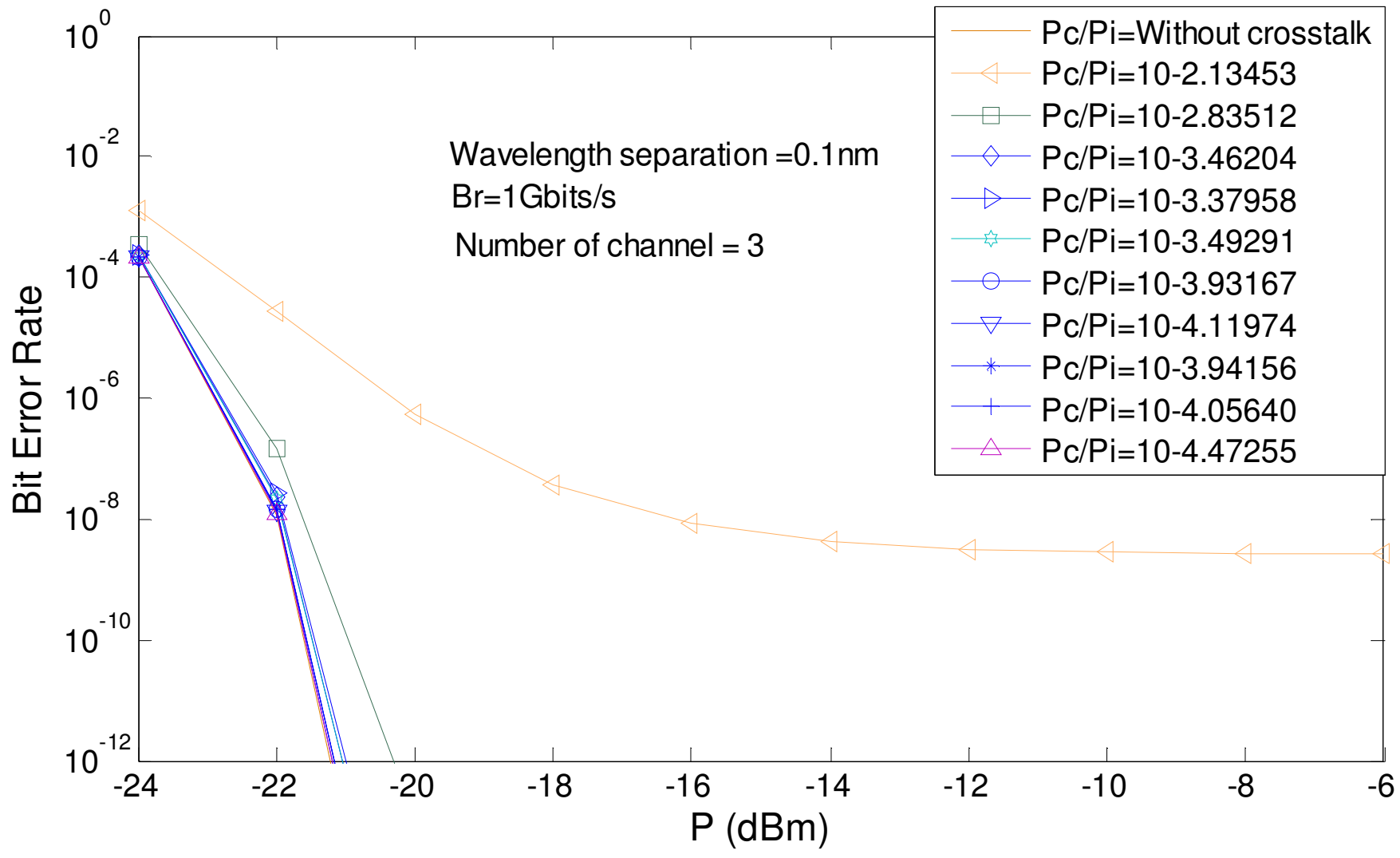

**Fig.7:** Plots of BER versus received power, Ps(dBm) for three channel crosstalk, bit rate, Br= 1Gbits/s

The bit error rate performance results for a WDM system with number of wavelength N=5 are depicted in Fig.8 with channel wavelength separation $\Delta\lambda$ as a parameter. It is noticed that the effect of crosstalk is more than that for N=5 as shown in Fig.7. As is evident from Fig.8 that BER floor occurs at around $10^{-6}$ when $\Delta\lambda$ =0.1 nm. Increasing the wavelength separation to 0.2 nm results in a drastic reduction in the value of BER due to reduced crosstalk. Similar results are also shown in Fig.9 for N=7 at 1Gbit/s bit rate.

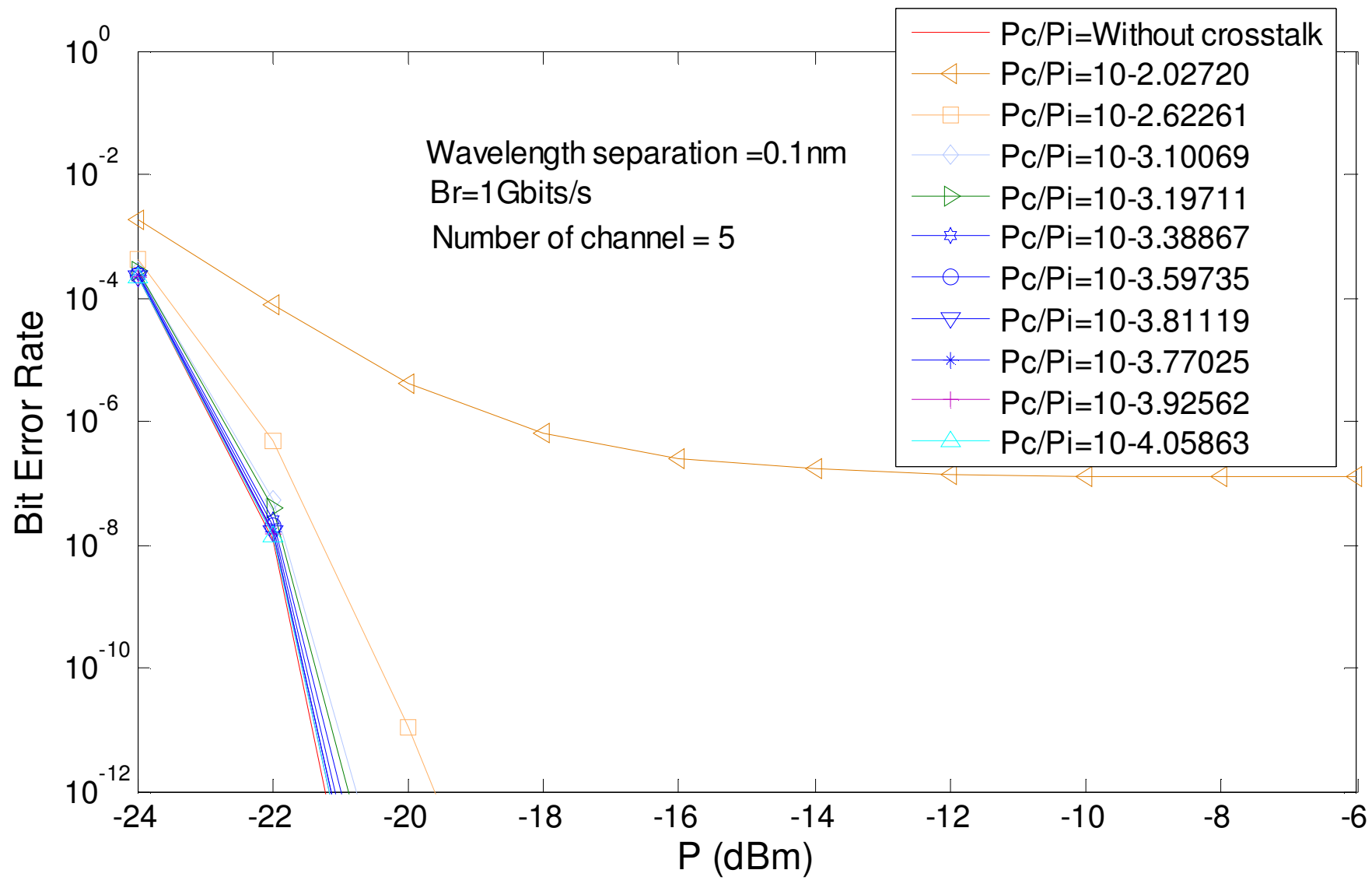

**Fig.8 :** Plots of BER versus received power, Ps(dBm) for five channel crosstalk, bit rate, Br= 1Gbits/s

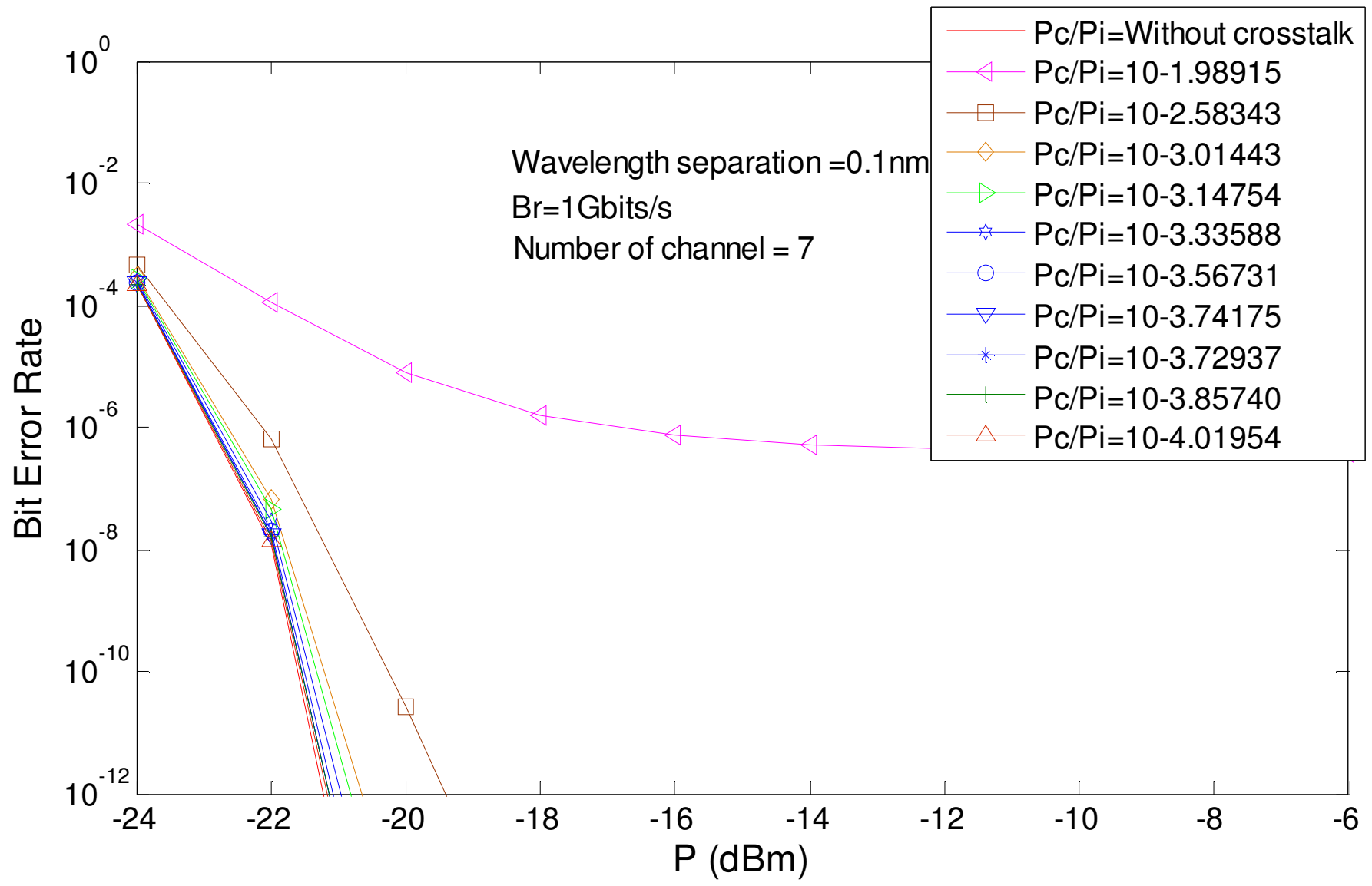

**Fig.9 :** Plots of BER versus received power, Ps(dBm) for seven channel crosstalk, bit rate, Br= 1Gbits/s

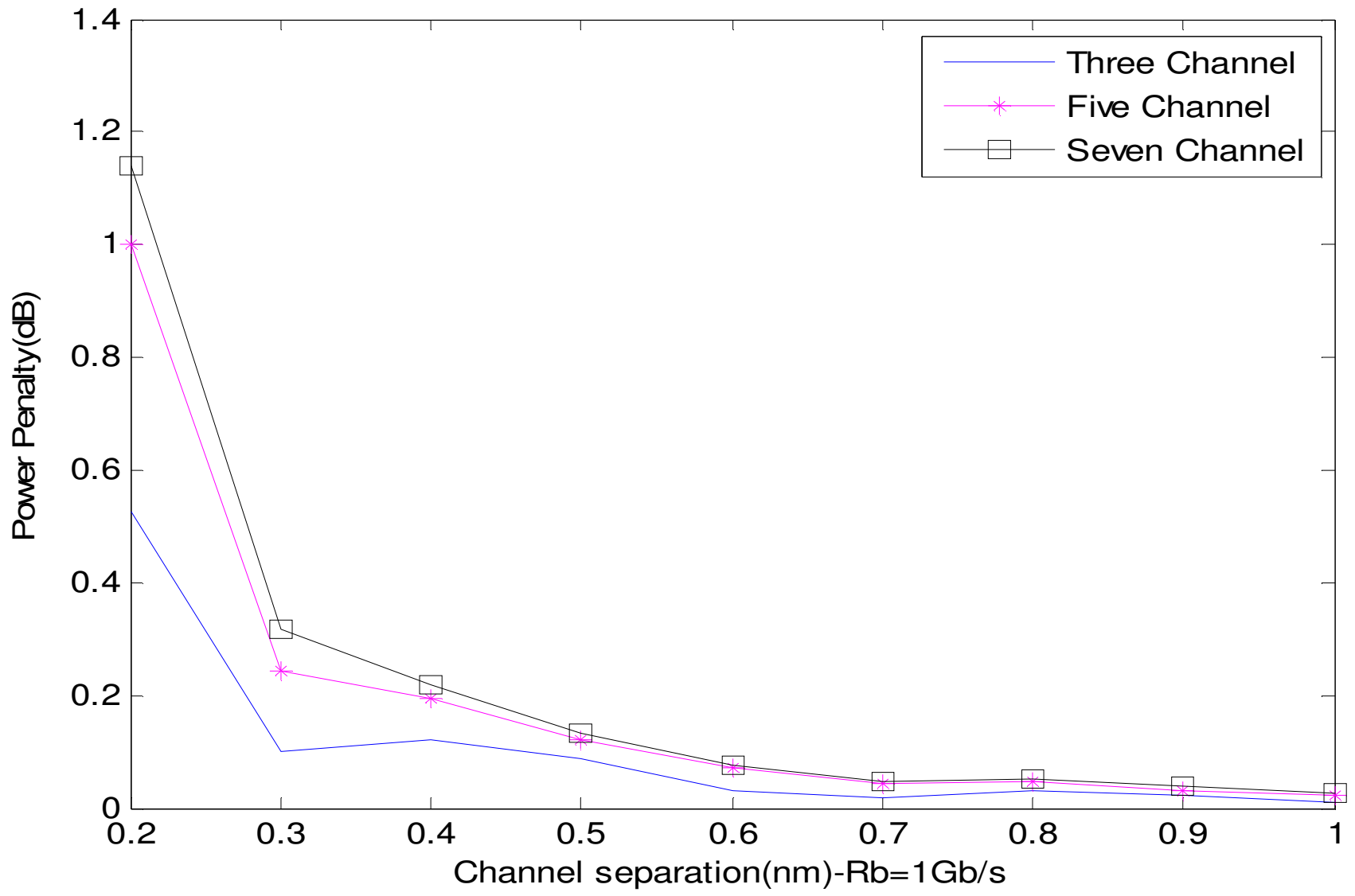

**Fig. 10:** plots of Power penalty versus channel separation for bit rate, Br= 1Gbits/s at BER = $10^{-9}$.

The increase in required received optical power in presence of OXC induced crosstalk at BER=$10^{-9}$ compared to the receiver sensitivity without crosstalk can be tuned as the power penalty due to OXC induced crosstalk. The plots of power penalty versus channel separation for bit rate, Br= 1Gbits/s at BER = $10^{-9}$ due to OXC induced crosstalk are depicted in Fig. 10. Plots are shown for three different values of WDM channels N=3, 5 and 7. Similarly we analyzed the whole system for another two different bit rate of 10Gbits/s and 20Gbits/s at BER = $10^{-9}$ due to OXC induced crosstalk are depicted in Fig. 11 and 12. Plots are also shown for three different values of WDM channels N=3, 5 and 7.

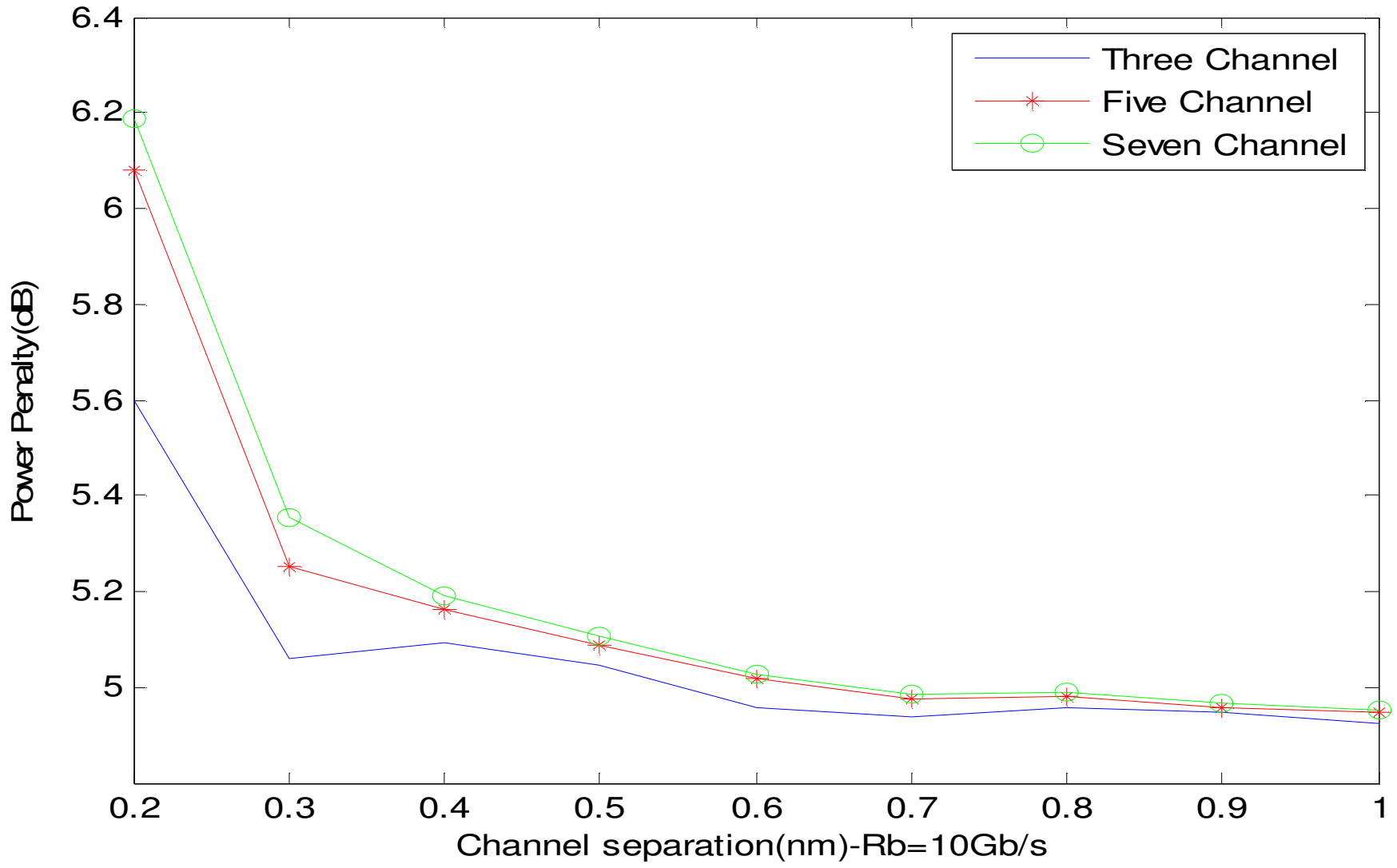

**Fig.11 :** Plots of power penalty versus channel separation for bit rate, Br= 10Gbits/s at BER = $10^{-9}$.

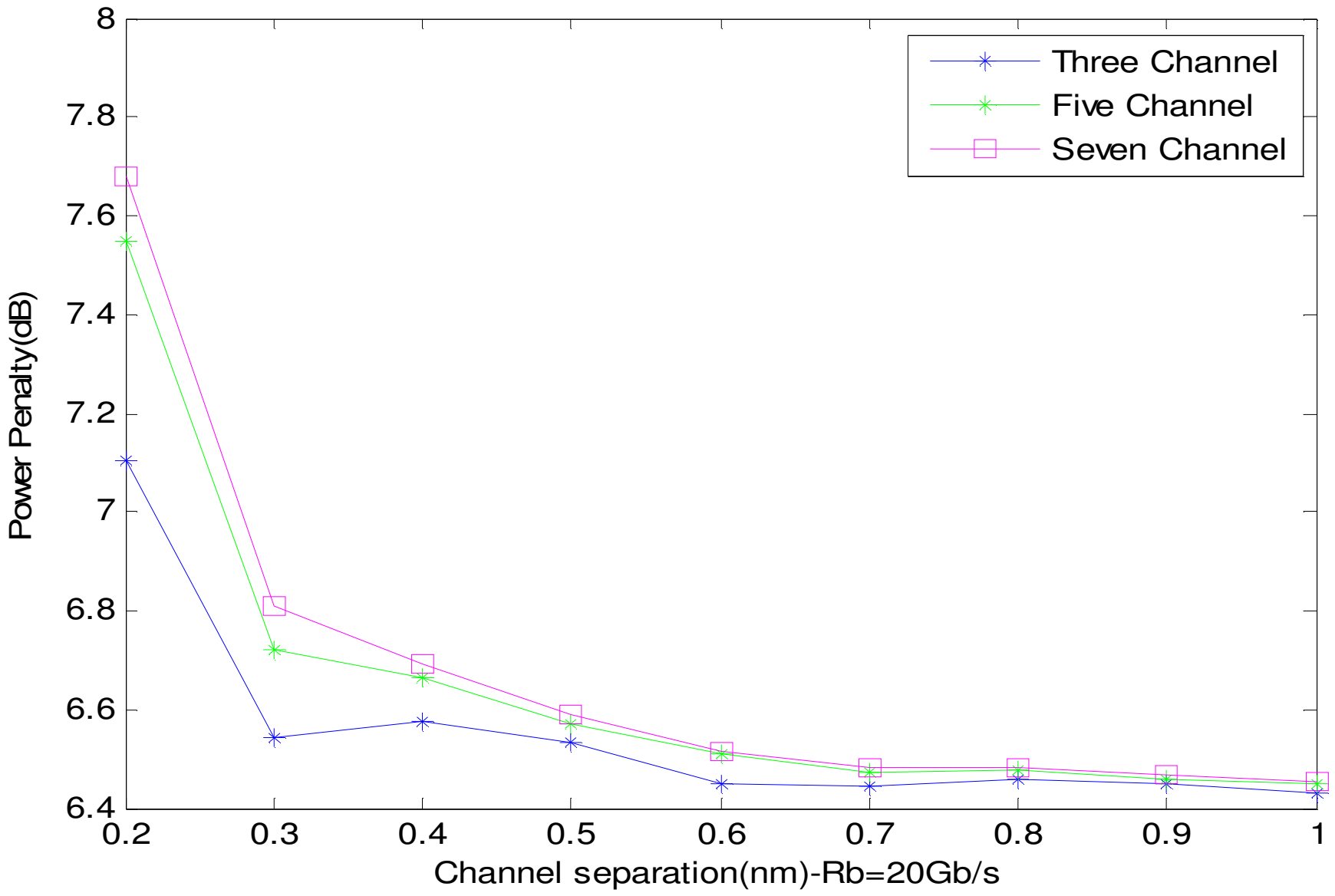

**Fig.12:** Plots of power penalty versus channel separation for bit rate, Br= 20Gbits/s at BER = $10^{-9}$.

From the overall analysis it is found that a high input power causes an extra penalty. Penalty also increases as we increase the number of input channels (N X M) i.e. the number of wavelengths and the number of input fibers but the penalty is more prominent with the number of input fibers than that of wavelengths at different bit rate.

# CONCLUSIONS

The results obtained here show the bit error rate performance of a WDM system considering the effect of crosstalk due to optical WDM MUX/ DMUX. In this paper the crosstalk sources have been identified and their total crosstalk is calculated based on analytical equations. Good qualitative agreements with the numerical simulations have been demonstrated. The power penalty due to crosstalk is evaluated at a BER of $10^{-9}$ for different transmission rate. The theoretical analysis shows that at high input power causes an extra penalty and penalty also increases as we increase the number of input channels. It also found that under different switching states, there exists a range of crosstalk performance. It shows that if the switch is improved then the number of fibers can be increased without significant penalty.

**Author**

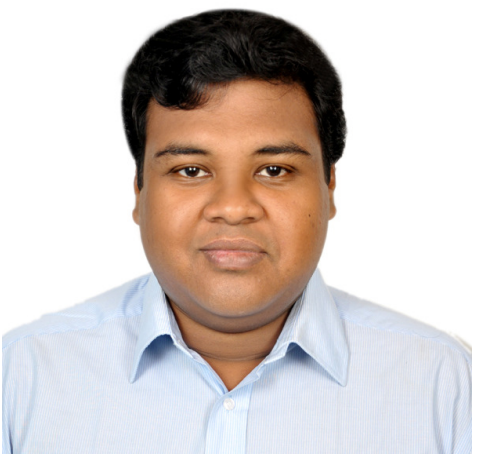

**Bobby Barua** received the B.Sc. and M.Sc. degrees in electrical and electronic engineering from Ahsanullah University of Science and Technology and Bangladesh University of Engineering and Technology respectively. He is currently Assistant Professor of Electrical and Electronic Engineering at Ahsanullah University of Science and Technology, Dhaka, Bangladesh. His research interests are in applications of information theory and coding to modern communication systems, specifically digital modulation and coding techniques for satellite channels, wireless networks, spread-spectrum technology, and space–time coding for multipath channels.